%% file: main.tex
\documentclass[runningheads,a4paper,10pt, final]{llncs}
\usepackage{graphicx}
\usepackage{hyperref}

\usepackage{microtype}

\usepackage{tikz}
\usepackage{amsmath}

\usepackage[misc]{ifsym}

\usepackage{csquotes}

\usepackage{adjustbox}

\usepackage{makecell}

\usepackage{colortbl}

\usepackage{multirow}

\usepackage{booktabs}

\usepackage{subcaption}
\usepackage{fp}

\usepackage{ifthen}

\newboolean{anonymous}
\setboolean{anonymous}{true} 
\newcommand{\anon}[1]{\ifthenelse{\boolean{anonymous}}{#1}{}}
\newcommand{\nonanon}[1]{\ifthenelse{\boolean{anonymous}}{}{#1}}
\newcommand{\ifanon}[2]{\ifthenelse{\boolean{anonymous}}{#1}{#2}}

\newcommand{\numAnnotIrrelevant}{6}
\newcommand{\percentageAnnotIrrelevant}{$1.5\%$}
\newcommand{\correctedPredPercentage}{$90\%$}
\newcommand{\numcccerts}{5394}
\newcommand{\refRichTotal}{1659}
\newcommand{\refRichPercentage}{$30.76\%$}
\newcommand{\numRefsTotal}{2712}

\newcommand{\NumPositiveReachWhenArchived}{$268$}

\newcommand{\refRichSmartcard}{1295}
\newcommand{\refRichSmartcardPercentage}{$74.08\%$}
\newcommand{\refRichSmartcardComp}{1080}
\newcommand{\refRichSmartcardCompPercentage}{$61.78\%$}
\newcommand{\refRichSmartcardPred}{390}
\newcommand{\refRichSmartcardPredPercentage}{$22.31\%$}

\newcommand{\refRichRelatedComp}{138}
\newcommand{\refRichRelatedCompPercentage}{$11.92\%$}
\newcommand{\refRichRelatedPred}{123}
\newcommand{\refRichRelatedPredPercentage}{$10.62\%$}
\newcommand{\NumRefsToArchivedOnIssuanceDate}{$40$}

\newcommand{\refRichOthersComp}{69}
\newcommand{\refRichOthersCompPercentage}{$2.77\%$}
\newcommand{\refRichOthersPred}{90}
\newcommand{\refRichOthersPredPercentage}{$3.62\%$}

\newcommand{\staleNL}{$1\%$}
\newcommand{\staleDE}{$23\%$}
\newcommand{\staleFR}{$19\%$}

\newcommand{\numManuallyAnnotated}{400}

\newcommand{\percentageMatch}{$82\%$}

\newcommand{\percentageMatchBinary}{$94\%$}

\newcommand{\sentFScoreMulticlass}{$0.79$}
\newcommand{\sentFScoreBinary}{$0.89$}

\newcommand{\numHyperparams}{12}

\usepackage[obeyFinal]{todonotes}

\usepackage{tcolorbox}

\begin{document}

\title{Chain of trust: Unraveling references among Common Criteria certified products}
\titlerunning{Unraveling references among Common Criteria certified products}

\author{Adam Janovsky\textsuperscript{\Letter}, {\L}ukasz Chmielewski, Petr Svenda, Jan Jancar, Vashek Matyas}
\institute{Masaryk University, Brno, Czechia \email{| adamjanovsky@mail.muni.cz}}
\authorrunning{A. Janovsky et al.}


\maketitle

\begin{abstract}
With \numcccerts{} security certificates of IT products and systems, the Common Criteria for Information Technology Security Evaluation have bred an ecosystem entangled with various kind of relations between the certified products. Yet, the prevalence and nature of dependencies among Common Criteria certified products remains largely unexplored. This study devises a novel method for building the graph of references among the Common Criteria certified products, determining the different contexts of references with a supervised machine-learning algorithm, and measuring how often the references constitute actual dependencies between the certified products. With the help of the resulting reference graph, this work identifies just a dozen of certified components that are relied on by at least 10\% of the whole ecosystem -- making them a prime target for malicious actors. The impact of their compromise is assessed and potentially problematic references to archived products are discussed.


\end{abstract}

\input{sections/introduction}
\input{sections/background}
\input{sections/related_work}

\input{sections/methodology}
\input{sections/results}

\input{sections/discussion}

\input{sections/conclusions}

\bibliographystyle{splncs04}
\bibliography{references.bib}

\end{document}

%% file: sections/introduction.tex
\section{Introduction}
\label{sec:intro}


Designing and budgeting secure and trusted systems would be a completely different world if security evaluations and certifications for computer systems were not around. The Common Criteria for Information Technology Security Evaluation (CC)~\cite{cc} constitute a pivotal element in the security evaluation and certification world for over quarter of a century. \numcccerts{} products\footnote{The data collected for this paper dates to November 1, 2023. We pledge to update the data for the camera-ready version of the paper.} have passed the CC certification process, with the majority incurring evaluation costs of hundreds of thousands of euros.

The plethora of certified products creates a complex network of interconnected components, frequently linking to other devices without clear reasons for these references. It remained unexplored until now whether the references indicate genuine dependencies.  In contrast, the dependencies between software packages across various programming languages have been extensively analyzed in various studies~\cite{decan2016topology,zimmermann2019small,liu2022demystifying}. This research has shed light on the level of trust placed in popular software libraries and their influence on other packages. By evaluating the extent of their impact, the community has been able to pinpoint packages with a significant reach, marking them as potential targets for malicious actors. This insight enables tighter monitoring of these critical parts, encouraging developers to critically assess and possibly reduce dependencies, thus navigating the fine line between code re-use and avoiding the dependency maze.

In our research, we explore the dependency-related security risks within the CC ecosystem by analyzing the references among certified products. In particular, we leverage the previously developed \texttt{sec-certs} tool~\cite{janovsky2023sec} to construct a directed reference graph, and we propose a supervised machine-learning method to annotate edges with reference meanings. Through this representation, we show how often the certified products rely on security functions of other devices, and we identify the high-value targets for adversaries and defenders, i.e., products with an extensive reach (many incoming references). Additionally, we conduct an empirical analysis to assess the impact that compromising such a highly referenced device would have on its surroundings. Furthermore, our investigation extends to identifying references that may pose risks, such as those pointing to outdated or archived products. We release the paper artifacts on GitHub~\cite{submission_repo}.

Our work has several important implications. First, upon discovering a vulnerability in a certified product, security analysts can swiftly identify other certified products that refer to it, considering the context of these references. Second, products with extensive reach warranting increased oversight can be put under tighter scrutiny, with our analysis providing data-backed support for such measures. Third, designers of new products can evaluate the trust associated with using certified products as sub-components, leading to better-informed decisions about dependencies. We frame our undertaking in the following research questions:

\begin{description}
    \item[RQ1:] \textit{What are the reasons for references among CC-certified products? How often do certificates reference each other across different product categories?}
    \item[RQ2:] \textit{What certificates have the highest reach? What would be the impact of their compromise, w.r.t. both direct and transitive references?}
    \item[RQ3:] \textit{How do referenced certificates age? Are there references from active to archived products, and how does the post-archival reach of products fade?}
\end{description}

\noindent
In this work, we deliver the following contributions:

\begin{itemize}
    \item We formalize the CC reference graph and develop a machine-learning method to automatically label its edges with codes signalling the reference context.
    \item We provide the first comprehensive analysis of inter-certificate references in the CC, concentrating on high-reach devices and aging products.
\end{itemize}




\noindent
\textbf{Paper roadmap}: The following section provides background on Common Crieria. In Section~\ref{sec:related_work}, we map the related work. Section~\ref{sec:methodology} explains our research methodology. We then present the results in Section~\ref{sec:results}. Section~\ref{sec:discussion} reviews the limitations and discusses the results further. We conclude in Section~\ref{sec:conclusions}.

%% file: sections/background.tex
\section{Common Criteria background}
\label{sec:background}

Common Criteria (CC)~\cite{cc} is an international standard (ISO/IEC 15408) for evaluating and certifying the security of IT products. The CC provides interoperability between numerous national certificate Authorizing Schemes (ASs), fostering trust in solutions certified under the Common Criteria Recognition Arrangement (CCRA)~\cite{assurance_continuity}. Within this agreement, the 
ASs base their certifications on evaluations conducted by accredited and independent security laboratories. 

In the CC standard~\cite{cc}, any security-related product can be certified, emphasizing the quality of the development process as a predictor of product security. The certification begins with the applicant submitting a Security Target (ST), which details the product's security specifications and goals. Applicants typically choose from a set of predefined Protection Profiles (PPs) that describe scenarios for use, like smartcards. An independent evaluator then checks if the product meets these specifications. The product achieves certification at one of Evaluation Assurance Levels (EALs)~\cite{cc_overview_2017}, ranging from 1 to 7. Higher EALs indicate more assurance in the security but also involve greater costs.

While there exist various categories of CC-certified products (from integrated circuits to databases), a large part of the CC portfolio concerns smartcards ($32\%$ of all certificates). Additionally, half of the currently active certificates are in levels considered to be relatively secure, with EAL4 or above. Within this secure segment, smartcards play a major role, making up 72\% of the certifications, and being certified (97\%) at EAL4 or higher.

Due to its broad applicability across various product categories, 
the CC scheme features diverse relationships between certificates.  Among these, composite evaluations~\cite{joint_composite} stand out by allowing multiple security products, both hardware and software, to be assessed together as a single layered system. This process mandates 
each component to be evaluated individually within a well-defined scope. A composite evaluation typically involves at least two parts: an \emph{underlying platform} and an \emph{application} that operates on this platform, with the platform typically being evaluated first. To enhance the security of composite products, certain national certification bodies, such as German BSI and the Dutch NSCIB, stipulate that the oldest certificate in a composite chain must undergo re-evaluation every 18 months.~\cite{BSI_highlevel,NSCIB_form}.

Other types of relationships between certificates are significant for our study. Firstly, several members of a product family might be certified individually while sharing some components. In this situation, a part of one evaluation can be reused for another one. Secondly, the relations between certificates issued at different times expressed via maintenance or re-evaluation reports reflect the evolution of security properties or scope updates over time.


%% file: sections/related_work.tex
\section{Related work}
\label{sec:related_work}
\noindent
\textbf{Quantitative analysis of the CC certification practice.} In the earlier work by Janovsky et al.~\cite{janovsky2023sec}, comprehensive processing of all CC certification artifacts into machine-readable documents was conducted, coupled with the pairing of certified products to CVEs from NIST's National Vulnerability Database. This effort allowed to measure the associations between security requirements and the vulnerabilities impacting the products. The authors presented case studies showcasing the utility of their tool in vulnerability assessment and mitigation. Nevertheless, critical aspects were deferred for future exploration. Although a method for constructing the vertices and edges of a reference graph was suggested, the graph was not formalized, or annotated with reference context that we provide.

To the best of our knowledge, only a handful papers on the CC ecosystem insights exist. Various statistics (limited to EAL4 products) were presented by Kaluvuri et al.~\cite{kaluvuri_quantitative_2014}. Furthermore, jtsec conducts regular scraping of CC artifacts~\cite{ccscraper-2022}, although their publications omit any reference analysis. 

\noindent
\textbf{Dependencies in software packages.} Initially probed by Decan et al. in~\cite{decan2016topology}, the package topologies between Python, R, and Javascript languages were compared. The key takeaway was that there are considerable differences between the treatment of package dependencies, each with distinct security implications. In 2019, Zimmermann et al.~\cite{zimmermann2019small} explored the dependencies in a densely connected Javascript ecosystem; a substantial portion of it relies on the security of just dozens of packages and their maintainers. The setting in the CC scheme is more favourable since the dependencies are often certified and thoroughly evaluated. Still, the emergence of a vulnerability may impact many other certified products, as was showcased by Janovsky et al.~\cite{janovsky2023sec} on several vulnerabilities like ROCA \cite{roca}. In the npm world, the extent of the vulnerability propagation was studied at scale~\cite{liu2022demystifying}, confirming that many packages depending (even transitively) on the vulnerable resource are often impacted. The recommended mitigation strategy for npm, to update dependencies, is hard to follow in the realm of certified products, where any significant update results in additional formal evaluation.

%% file: sections/methodology.tex
\section{Methodology}
\label{sec:methodology}

We start this section by introducing the notion of the CC reference graph. We then show how we constructed the vertices and edges of the graph from the certification artifacts and how we identified the different contexts of inter-certificate references. Finally, we illustrate how we built and evaluated a model to learn the reference meanings from the certification artifacts. In analogy to research on software packages~\cite{zimmermann2019small}, we formally define the \textit{Common Criteria reference graph}.
\vspace{3mm}
\newline
\framebox{
    \begin{minipage}{0.96\textwidth}
    \noindent
     Denote the set of all CC-certified products as $\mathcal C$. The CC reference graph is a pair $(\mathcal C, E)$ where an edge $(c_i, c_j) \in \mathcal{C} \times \mathcal{C}$ (for $i \neq j$) belongs to $E$ iff the product $c_i$ references the product $c_j$ inside its certification artifacts.
    \end{minipage}
}

\subsection{Building the vertices and edges}
Following the approach of Janovsky et al.~\cite{janovsky2023sec}, we use the public CC artifacts in this study, typically available in the PDF format. These include a variety of documents: a \emph{security target} (specification of the certified product), \emph{web pages with additional metadata and summaries} generated by the CC portal, a \emph{certification report} (a summary of the certification results), a \emph{maintenance report} (description of minor changes in an already certified product), and a \emph{protection profile}. 

The~\texttt{sec-certs} tool~\cite{janovsky2023sec} collects certification artifacts and features for all certified products and one can use an API to interact with the tool. Among others, a unique identifier (certificate id) is collected for each certified product. Additionally, the presence of identifiers of different certificates is also collected for each of the certified products with high precision (99\%). We build the reference graph from these attributes, i.e., we include the edge $(c_i, c_j)$ into the graph iff either the certification report or the security target of $c_i$ explicitly mentions\footnote{The \texttt{sec-certs} tool monitors for such mentions with 25 distinct regular expressions described in the related study~\cite{janovsky2023sec}.} the certificate id of $c_j$. To this end, the complete dataset contains \numRefsTotal{} references (edges) in \numcccerts{} products (vertices).

\subsection{Categorization of reference meanings}

To capture the reference context of the edge $(c_i, c_j)$, we introduce the edge-labelling function $\ell$, which maps each edge to its categorical code. Before showing how we automatically learn the function $\ell$ on the whole dataset of \numRefsTotal{} edges, we describe how we deduced the different reference contexts. 

To determine the initial codes, we exploratively studied the certification artifacts. First, one co-author randomly sampled 100 different edges $(c_i, c_j)$ from the established graph and went through the text artifacts of $c_i$ to develop the initial codes (i.e., category types) of reference meanings. These codes were then observed by a second co-author who went through a different set of 100 edges. The final categories were then refined in a discussion between these two co-authors, during which the descriptions and examples for codes were synthesized. Overall, we discovered two fundamental reasons (binary codes) that necessitate a reference:

\begin{itemize}
    \item There is some \textbf{Component reuse (C)} between the referenced and referencing products.
    \item The referenced product is a \textbf{predecessor (P)} of the referencing device.
\end{itemize}

This taxonomy can be further refined to capture more fine-grained context\footnote{We provide the complete codebook for both the coarse-grained and the fine-grained categories from our replication package~\cite{submission_repo}.}, yet at the expense of subtle differences between the codes that are challenging to be automatically extracted from the certification artifacts due to ambiguous natural language used. 
We provide this finer taxonomy mostly to describe the referencing culture, and we work with the binary codes when presenting results in the next section. The fine categories (denoted later as multiclass codes) are:

\begin{itemize}
    \item \textbf{Component used (C)}: The referenced product is used as a whole in the referencing product (e.g., smartcard referencing the underlying IC).
    \item \textbf{Component shared (C)}: The referenced product shares some components with the referencing product. 
    \item \textbf{Evaluation reused (C)}: The evaluation results of the referenced product were used for evaluation of the referencing product due to reasons that the annotators could not resolve even after manual inspection.
    \item \textbf{Re-evaluation (P)}: The referencing product is a formal re-evaluation of the referenced product. For the exact definition of re-evaluation, see assurance continuity requirements~\cite{assurance_continuity}.
    \item \textbf{Previous version (P)}: The referenced product is a previous version of the referencing product and re-evaluation is not explicitly mentioned.
\end{itemize}

Having the codebook available, we sampled \numManuallyAnnotated{} edges from the reference graph. Two co-authors then independently reviewed the certification artifacts for these instances and manually annotated the edges with codes. A special code \textit{unknown} was used when the annotator was unable to decide on the right category.
During the annotation task, we noticed \numAnnotIrrelevant{} (\percentageAnnotIrrelevant{}) cases where the reference was irrelevant to the studied artifacts, e.g., due to a typo in the referenced id or the reference being mistakenly left out in the document. Given the number of such cases, we further disregard it in our work. The annotation agreement for the 5-class taxonomy between the co-authors after the first annotation round was \percentageMatch{}. Even better, the agreement on the binary codes was \percentageMatchBinary{}. The conflicting and \textit{unknown} instances were then resolved in a discussion between the annotators. We are now ready to define \textit{certificate reach} as follows:
\vspace{3mm}
\newline
\framebox{
    \begin{minipage}{0.96\textwidth}
    \noindent
        For every certificate $c \in \mathcal C$, we define its certificate reach as the number of certified products from which a path to $c$ exists in the reference graph, such that all edges on the path are labelled with the \textit{component-reuse} (C) code. 
    \end{minipage}
}

\subsection{Learning the edge-labelling function}

\begin{figure*}
    \captionsetup{font=footnotesize}
    \centering
    \includegraphics[width=1\textwidth]{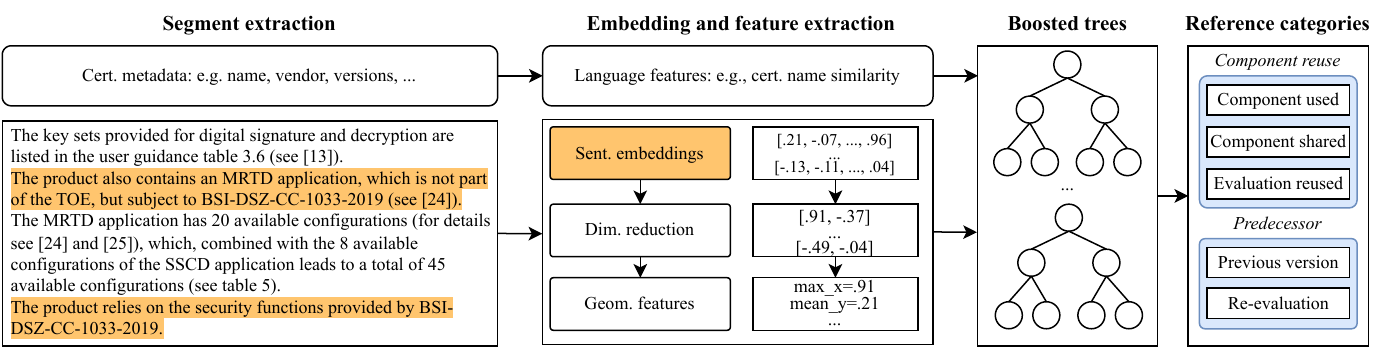}
    \caption{A high-level overview of the edge-labelling model.}
    \label{fig:methodology:framework}
\end{figure*}

The reference context for $(c_i, c_j)$ can often be reliably inferred from the sentences surrounding the identifier of the referenced certificate $c_j$ in the certification artifacts of $c_i$. For instance, the sentence \textit{\enquote{this is a re-certification of BSI-123-456-CC}} clearly signals the \emph{re-evaluation} category. To learn the function $\ell$, we first extract the sentences surrounding the reference identifiers from the documents. After feature extraction, we train a machine learning classifier to map the features to the final categories. This process is illustrated in Figure~\ref{fig:methodology:framework} and we describe it closely in what follows. We split the \numManuallyAnnotated{} manually annotated edges in a 50:50 fashion into train:evaluation sets.

\subsubsection{Segment extraction.} From the certification report and the security target documents, we extract the sentence containing the reference id together with 2 preceding and 1 succeeding sentences\footnote{The ideal surrounding length was identified through a hyperparameter search.}. For each edge, we extract one or more of such \textit{reference segments} that likely carry the context of the reference.  
    
\subsubsection{Vector embeddings and feature extraction.} We then encode each reference segment with a sentence transformer into a long vector of floats. As a baseline, we use TF-IDF vectors instead of embeddings. We then reduce each vector to 2 dimensions (results are superior to 3D or more dimensions) with UMAP~\cite{mcinnes2018umap} and PCA.  Since one or more vectors were extracted for each edge, we aggregate the results with various statistical features: median, max, variance and so forth (their complete list is available from the replication package~\cite{submission_repo}). Aside from the embeddings, we also sample various language features directly from the certificate metadata. For example, we measure the title similarity for the referencing and referenced product (high similarity suggests predecessor context).  This process yields a single feature vector for every edge.

\subsubsection{Classifier training and evaluation.} We use gradient-boosted trees to build a classifier mapping the edges to the reference categories. In each stage of our pipeline, we identified the hyperparameters influencing the model performance and estimated the initial values. The hyperparameter tuning was done for each stage separately, fixing the parameters from the other stages and finding the best values for the examined stage. Overall, we finetuned \numHyperparams{} hyperparameters using 5-fold cross-validation; we list those in the replication package~\cite{submission_repo}. To evaluate the model, we used a weighted F1 score as a target metric. We evaluated three different model variants: \emph{(i)} A random guess,  \emph{(ii)} boosted trees with TF-IDF, and  \emph{(iii)} boosted trees with sentence embeddings. For each of those models, we evaluated both multiclass and binary categories. 

\begin{figure}
  \centering
  \begin{minipage}{0.58\textwidth}
    \centering
    \adjustbox{valign=t}{%
    \begin{tabular}{l|c|c}
        \toprule
        \textbf{Model} & \multicolumn{2}{c}{\textbf{Weighted F1 score}} \\
         & {Binary } & Multiclass\\
        \midrule
        Sent. transformers & \textit{0.89} & \textit{0.79} \\
        TF-IDF &  0.87 & 0.77 \\
        Random guess & 0.65 & 0.49 \\
        \bottomrule
    \end{tabular}%
    }
    \captionof{table}{\label{tab:model_evaluation}Weighted F1 scores for different classifiers. Both binary and fine-grained taxonomies were evaluated.}
  \end{minipage}
  \hfill
  \begin{minipage}{0.38\textwidth}
    \centering
    \adjustbox{valign=t}{\includegraphics[width=1\textwidth]{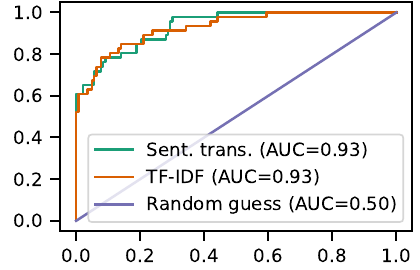}}
    \captionof{figure}{\label{fig:roc}Receiver operating characteristic of different classifiers for the binary taxonomy.}
    
  \end{minipage}
\end{figure}

The summary of the classifier performance is depicted in Table~\ref{tab:model_evaluation}, and the receiver operating characteristic in Figure~\ref{fig:roc}. Overall, the sentence transformers are superior with  F1 score \sentFScoreBinary{} (\sentFScoreMulticlass{} on multiclass), having an edge over the TF-IDF method ($0.87$ / $0.77$). Our classifier also beats the inter-annotator agreement of human experts and dramatically improves over the random guess augmented by knowledge of class imbalance. 
Even better, when incorporating the 394 relevant samples annotated by two co-authors as error-free annotations, the expected accuracy is over \correctedPredPercentage{} on the complete dataset. Such a classifier is a solid base for automatically classifying references in future certificates.

%% file: sections/results.tex
\section{Reference analysis}
\label{sec:results}

We now leverage the obtained reference graph with annotated edges (described in Section \ref{sec:methodology}) to answer the proposed research questions using data analysis.

\subsection{Referencing culture (RQ1)}

\begin{figure}[t]
\centering
\subcaptionbox{Reference context freq.\label{fig:subfig:context_prev}}
{\includegraphics[width=0.32\textwidth]{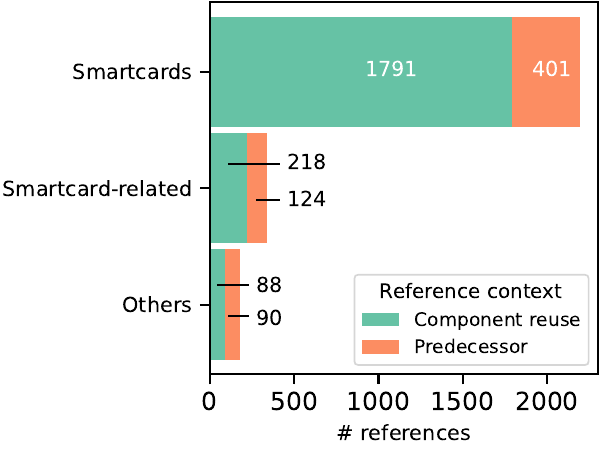}}
\subcaptionbox{Avg. \# trans. refs\label{fig:subfig:timeplot_avg_refs}}
{\includegraphics[width=0.32\textwidth]{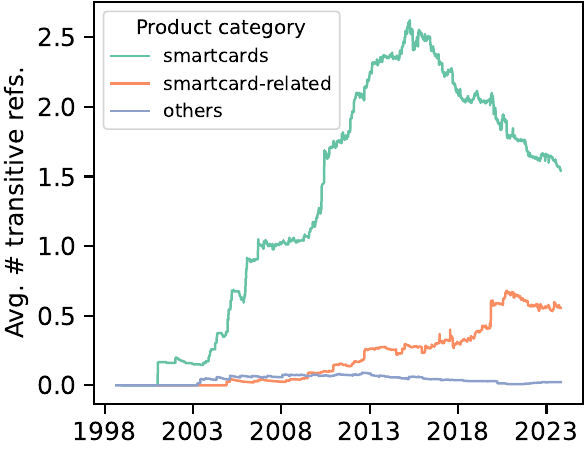}}
\subcaptionbox{Average product reach\label{fig:subfig:timeplot_reach}}
{\includegraphics[width=0.32\textwidth]{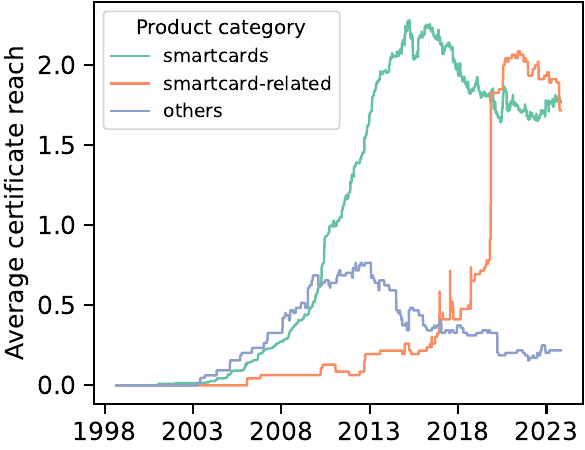}}
\caption{Subplot~(\subref{fig:subfig:context_prev}) displays the reference context popularity among categories. In subplot~(\subref{fig:subfig:timeplot_avg_refs}), the evolution of the average number of transitive \emph{component-reuse} references in time is depicted. Similarly, subplot~(\subref{fig:subfig:timeplot_reach}) shows how the average product reach evolves.}\label{fig:time_evolution}
\end{figure}

Out of the total \numcccerts{} products, \refRichTotal{} (\refRichPercentage{}) have at least one reference. However, there is a notable contrast in the referencing habits among smartcards (ICs, Smart Cards and Smart Card-Related Devices and Systems category), smartcard-related devices (Trusted Computing, Products for Digital Signatures, Other products and devices categories), and products from different categories. As illustrated in Figure~\ref{fig:subfig:context_prev}, majority (\refRichSmartcard{}, \refRichSmartcardPercentage{}) of smartcards reference some other product. Specifically, \refRichSmartcardComp{} (\refRichSmartcardCompPercentage{}) smartcards do reference some other certified product in a \emph{component-reuse} relation, while \refRichSmartcardPred{} (\refRichSmartcardPredPercentage{}) smartcards reference some predecessor. The smartcard-related products are much less connected: \refRichRelatedComp{} (\refRichRelatedCompPercentage{}) have a component-reuse reference, while \refRichRelatedPred{} (\refRichRelatedPredPercentage{}) reference their predecessor. The devices unrelated to smartcards very rarely utilize references, with only \refRichOthersPred{} (\refRichOthersPredPercentage{}) products referencing some predecessor and \refRichOthersComp{} (\refRichOthersCompPercentage{}) products engaging in component-reuse.

To determine whether the average vulnerability impact aggravates in time, we analyzed the reference graph temporal evolution. According to Figure~\ref{fig:subfig:timeplot_avg_refs}, smartcards reached their highest point with an average of 2.5 transitive component-reuse references in 2017. By November 2023, an average smartcard relies on the security functions of nearly two other certified products. Furthermore, Figure~\ref{fig:subfig:timeplot_reach}, shows that as of 2023, both smartcards and related devices extend their reach to almost two additional products. It is also apparent that the limited referencing among other categories demonstrates a consistent pattern. In summary, our findings regarding the referencing culture are as follows:
\vspace{3mm}
\newline
\framebox{
    \begin{minipage}{0.96\textwidth}
    \noindent
        The primary motives for cross-referencing among CC certificates are \emph{(i)} component reuse among devices and \emph{(ii)} reference to a predecessor. Smartcard products particularly favour component reuse, with the majority depending on at least one other certified product. The reach of an average smartcard has incrementally increased to approximately 2, around which it currently stabilizes. Although smartcard-related products occasionally have certified dependencies, products from other categories remain largely isolated.
    \end{minipage}
}

\subsection{Certified products with high reach (RQ2)}

Within our reference graph induced by component-reuse edges, we examined ten products with the highest reach. The champion, \href{https://seccerts.org/cc/dd000f356b48904d}{BSI-DSZ-CC-0813-2012} (Infineon smartcard IC M7794 A2 with ROCA-vulnerable RSALib v1.02.013), is transitively referenced from 77 other products. Combined, the top-10 transitively referenced products influence 196 other devices. The share of all active smartcards relying on the top-10 products has varied over time but has consistently exceeded 10\% since 2004. By November 2023, the current top-10 products affected 23\% of all valid smartcards. With such influence, this makes the high-reach products  extremely attractive targets for adversaries. Notably, all of the top-10 devices are, in fact, microcontrollers or some integrated circuits, often with cryptographic libraries. Our analysis revealed that the higher reach of the product is positively associated with a higher Evaluation Assurance Level, and hence the trust in that device. This was confirmed with Spearman's rank correlation test with the resulting association of $0.23$ (p-value  $<2.73\textrm{e}^{-23}$, one-sided alternative).

To better understand the threat that a vulnerability in a high-reach device could have on its surroundings, we measured to how many products it may propagate. In the component-reuse sub-graph, we selected all weakly-connected components with at least 10 certified products (15 different components in total)\footnote{The largest component, an outlier with 707 certified products, was infeasible to be analyzed manually and was excluded from the analysis.}. In each of the components, we monitored the product with the highest reach and manually annotated all its incoming transitive references with labels from the fine-grained categorization. We then counted the number of \emph{component used} references in these trees. This is because a critical vulnerability in a node most likely propagates through the component-used edge, as supported by the previous case studies~\cite{janovsky2023sec}. The high-reach nodes were all microcontrollers or ICs, with the exception of one operating system. We conclude that in the majority of the studied components, a critical flaw might spread dramatically to the derived products. In total, we labelled 170/245 (69\%) edges with component-used code. On (macro-)average, 70\% of the whole graph component is susceptible to the vulnerability originating from the wide-reach product. All studied components, the labelled edges and a summary table are available from the replication package~\cite{submission_repo}. This allows us to answer RQ2:
\vspace{3mm}
\newline
\framebox{
    \begin{minipage}{0.96\textwidth}
    \noindent
        Just a dozen of smartcard devices influence more than 10\% of the whole ecosystem at any given time. The typical high-reach device is an integrated circuit at the bottom of the smartcard component stack, implementing critical cryptographic functionality. These devices are generally evaluated to high assurance levels, EAL5+ or more.
        Our experiment showed that such high-reach devices are indeed used as components in nearly 70\% of the products that reference them. As a consequence, critical flaws in these high-reach devices would likely spread to many other products, crippling the broader surrounding. 
    \end{minipage}
}


\subsection{Ageing of referenced products (RQ3)}
\label{sec:subsec:rq3}

CC certificates have a 5-year validity according to operational guidelines~\cite{ccdb_2021}, with some schemes (e.g., German and Dutch~\cite{BSI_highlevel,NSCIB_form}) requiring a re-evaluation of the products in a composite chain every 18 months to mitigate risks from emerging threats. In this sub-section, we analyze whether archived certificates are being referenced and we check whether the national schemes follow the 18-month policy. 

We identified \NumRefsToArchivedOnIssuanceDate{} products that referenced an \emph{archived} component \emph{when they were issued}. A manual review showed that, in 17 instances, the certificate archival date was incorrectly marked as its issuance date in the Common Criteria portal, while the dates in the certification reports were correct. We also confirmed this in correspondence with one vendor of such certificates.
Furthermore, 6 products do reference components archived 5-12 months earlier, indicating these referenced products were likely active but nearing archival during the evaluation period. 

Eight certificates for various versions of the KoCoBox MED+ Netzkonnektor 
use the archived \href{https://seccerts.org/cc/d9bfffa3cc6d1c53/}{STARCOS 3.6 COS C1} smartcard, with the latest versions issued in June 2023, almost three years after the 
archival. This raises questions about whether documentation was not updated with newer references or if certain schemes and labs overlook the use of archived products. Additionally, four certificates were collectively archived following a vulnerability discovery, yet their evaluation outcomes were leveraged to quickly issue new certificates referencing those recently archived. Remarkably, the \href{https://seccerts.org/cc/83132bcbd3e7572c/}{CombICAO Applet v3 ePassport}, released after these re-certifications, still cites the compromised components in its certificate 
but omits them from the security target. This could indicate documentation discrepancies or even a \enquote{race condition} in the certification process, where the vulnerability in the underlying product was disclosed after the assurance evaluation yet before the final certificate was issued. 

Additionally, 4 references actually represented a predecessor relationship but were misclassified as component reuse by our model. In another instance, a vendor failed to update a security target with a new reference, leading to a false positive alert, although the corresponding certification report was accurate.

\begin{figure}[t]
\centering
\subcaptionbox{Ratio of component-reuse referenced certificates with $>0$ reach at $n$ days post-archival (only includes products with positive reach on the date of their archival).\label{fig:subfig:cdf_half_life}}
{\includegraphics[width=0.45\textwidth]{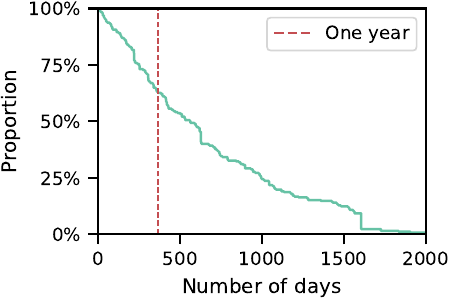}}
\hfill
\subcaptionbox{CDF: the age of the referenced certificate on the issuance date of the referencing certificate.\label{fig:subfig:ref_comp_age}}
{\includegraphics[width=0.45\textwidth]{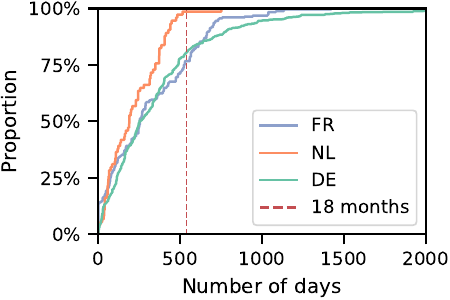}}
\caption{Analysis of referenced certificates with respect to archiving and ageing.}
\label{fig:halftime_archival}
\end{figure}

We also explored the rate at which the reach of certified products fades to zero following their archival, when considering the references from active products. We focused on \NumPositiveReachWhenArchived{} products that maintained a positive reach at the time of their archival. The findings reveal that the transition to zero reach extends beyond a year on average, as depicted in Figure~\ref{fig:subfig:cdf_half_life}.


We extended our analysis to still active but ageing products; specifically, we examined the references to certificates older than 18 months to check for adherence to the 18-month validity policy upon re-evaluation. In Figure~\ref{fig:subfig:ref_comp_age}, we illustrate the Cumulative Distribution Function (CDF) for the age of certificates referenced in smartcard composite evaluations across major schemes: German, French, and Dutch. Across these schemes, NL complies with their policy in nearly all products (\staleNL{} violate this rule), while FR and DE lag, with \staleFR{} and \staleDE{}, violating the rule respectively\footnote{The average for the whole CC is close to the numbers of DE and FR. The most notable exception is Spain, with only 43\% products complying with the 18-month policy. Moreover, recently issued FR products (2020 and newer) improved w.r.t. following this policy, with only 4\% violating the rule. DE does not show such a trend.}. We do not have a clear explanation for this discrepancy. While our analysis might be slightly affected by the noise from some re-assessments not being published, we believe this analysis shows an interesting trend differentiating the national CC schemes. Overall, we answer RQ3 as follows:
\vspace{3mm}
\newline
\framebox{
    \begin{minipage}{0.96\textwidth}
    \noindent
        In the CC ecosystem, referencing archived certificates is rare: by November 2023, only 14 different products were found to use components from archived certificates at the time of issuance.
        The persistence of archived products with positive reach, however, extends well beyond a year. 
        We observed that the Dutch scheme is the strictest towards old components in composite evaluations. 
        In contrast to this, some evaluation bodies certify products that include components nearing the archival or those already archived.
    \end{minipage}
}

%% file: sections/discussion.tex
\section{Discussion}
\label{sec:discussion}

Our study works with certification documents that have been generated by humans, with the prevailing expectations of their authors that these documents will be utilized by humans. Structure of these documents have naturally evolved over the past 25+ years. A greater availability of metadata related to the certification process would simplify the analysis and enhance the certification transparency. Let us delve into most noteworthy observations and issues.

\subsubsection{Variability in CC reference patterns.}
The use of references in CC varies across product types and certification bodies. Smartcards stand out for their extensive use of references. We argue that such dense reference network enhances security through independent evaluation of sub-components and the ability to track vulnerability spread via explicit, transient dependencies. 
The frequent use of references in smartcards can be attributed to the nature of these devices, which are built from multiple certified layers to facilitate reuse and reduce costs. This is a marked difference from, e.g., operating systems, which, despite their greater code volume and complexity, seldom rely on certified dependencies.

\subsubsection{References represent genuine dependencies.}
Programming language package managers typically come with an explicit dependency graph. No such thing is available for the CC, and our investigation aimed to fill this gap. Our research found that vendors are strongly motivated to explicitly reference other certified products, whether as sub-components or predecessors. This approach allows them merely to prove compliant use during evaluation rather than evaluate the security of the referenced products. Consequently, almost all references indicate a strong link between products, with a minimal number of irrelevant instances. In the CC, references can be tracked using identifiers in the certification documents; a method confirmed effective in previous research~\cite{janovsky2023sec}. Nonetheless, the occasionally ambiguous terminology surrounding the reference identifiers limited our ability to distinguish only between component reuse and predecessor contexts. Despite an approximate 10\% error rate of our model (given to a large extent by inherent noise in the certification documents and missing reliable metadata), we believe our analysis reliably identifies overall trends. We also developed a fine-grained taxonomy of reference contexts, though more suited for manual annotation tasks.

\subsubsection{Strategic scope reductions.}
Recall from Section~\ref{sec:subsec:rq3} that we observed instances where vendors opted to narrow the certification scope following the discovery of vulnerabilities, rather than addressing the issues and pursuing re-certification. While some situations justify this approach -- such as components that cannot be updated -- in general, such practice raises concerns and merits further investigation in future research.  We also noted how the demands of the certification process can influence the shape of the final product. To meet the certification requirements, products may adopt a more strict development approach, incorporating additional safeguards not found in non-certified counterparts. Yet, this adherence can also prompt vendors to deliberately limit the evaluation scope of their product, excluding essential features from the certification scope. Vendors might exclude certain functionalities from their certification claims, despite these features being centric to the nature of the certified product. An illustrative case is the certified \emph{network} camera (certificate \href{https://seccerts.org/cc/16ba6dab2c5c4b13/}{SERTIT-115}), which assumes a secure network interface. This anomaly in the certification landscape deserves a broader discussion about the motivation and implications of such practices.

\subsubsection{Exploring potential vulnerability spread through references.}
The presence of a component-reuse reference is a strong indicator of vulnerability transmission across the graph edge, yet it does not imply it. Accurately assessing this risk requires a deep understanding of the certified configuration scope, including what components are employed and what functionality is not used. The boundaries of the certified products are communicated through complex language in the certification documents, and their automatic identification remains an open problem -- we merely point out this issue to be worth considering for future work with a significant expert-driven manual analyses. Our investigation was partly inspired by the notable ROCA vulnerability~\cite{roca} and its impact on numerous certified products~\cite{janovsky2023sec}. Our goal was to determine the likelihood of similar incidents occurring, considering their origin in products with a high reach. Also, while direct references in composite products are typically highly relevant, the relevance of indirect references is less certain, as they may utilize less functionality from the products they reference.

%% file: sections/conclusions.tex
\section{Conclusions}
\label{sec:conclusions}

We provided what we believe is the first wide examination of dependencies between CC-certified products, achieved through a systematic study of the inter-certificate references. No analysis over the certification documents available today can determine and completely correctly classify all references within these documents. We proposed a supervised machine-learning algorithm to extract the reference context, further leveraged in our CC reference graph. This fully automated method extends also to newly released certificates. We published the relevant source code and paper artifacts on GitHub~\cite{submission_repo}.

We showed that dependencies are especially favoured in smartcards, revealing that over 10\% of all CC-certified products rely on the security functions delivered by just a dozen integrated circuits and microcontrollers. As was previously illustrated by the ROCA vulnerability, a critical flaw might spread to many certified products~\cite{janovsky2023sec,roca}. By modelling the vulnerability propagation, we provided the evidence that this was no outlier but rather an expected scenario, with more such events in the store. Further, we demonstrated that some products rely on archived products and that the reach of archived products declines slowly. 

Our study provides critical insights that enable CC stakeholders to make well-informed choices about product dependencies and to enforce robust security measures for prominent components with an extensive reach. We hope that our study will contribute to a shift in the way certification documents
are written, moving from the \enquote{produced by humans and consumed by humans} paradigm to
more accurate computer-assisted naming, indexing, and to metadata production.

\section{Acknowledgements}

This preprint has not undergone any post-submission improvements or corrections. The Version of Record of this contribution is published in ICT Systems Security and Privacy Protection. SEC 2024. IFIP Advances in Information and Communication Technology, and is available online at \url{http://dx.doi.org/10.1007/978-3-031-65175-5_14}.